\documentclass[10pt,letterpaper]{article}
\usepackage{amssymb}
\usepackage{amsmath}
\usepackage{graphicx}
\usepackage{bm}
\usepackage{cite}
\usepackage{fullpage} 

\begin{document}
\title{Photonic circuits for generating modal, spectral,
and polarization entanglement}
\maketitle
\hyphenation{wave-guide wave-guides}

\begin{center}
\author{Mohammed~F.~Saleh,$^{1}$ Giovanni~Di~Giuseppe,$^{2,3}$ Bahaa~E.~A.~Saleh,$^{1,2}$
and Malvin~Carl~Teich,$^{1,4,5}$}
\textit{$^1$Quantum Photonics Laboratory, Department of Electrical \& Computer Engineering, Boston University, Boston, MA 02215 USA\\
$^2$Quantum Photonics Laboratory, College of Optics and Photonics (CREOL), University of Central Florida, Orlando, FL 32816 USA\\
$^3$School of Science and Technology, Physics Division, University of Camerino, 62032 Camerino (MC), Italy\\
$^4$Department of Physics, Boston University, Boston, MA 02215 USA\\
$^5$Department of Electrical Engineering, Columbia University, New
York, NY 10027 USA}
\end{center}

\begin{abstract}
We consider the design of photonic circuits that make use of
Ti:LiNbO$_{3}$ diffused channel waveguides for generating photons
with various combinations of modal, spectral, and polarization
entanglement. Down-converted photon pairs are generated via
spontaneous optical parametric down-conversion (SPDC) in a two-mode
waveguide. We study a class of photonic circuits comprising: 1) a
nonlinear periodically poled two-mode waveguide structure, 2) a set
of single-mode and two-mode waveguide-based couplers arranged in
such a way that they suitably separate the three photons comprising
the SPDC process, and, for some applications, 3) a holographic Bragg
grating that acts as a dichroic reflector. The first circuit
produces two frequency-degenerate down-converted photons, each with
even spatial parity, in two separate single-mode waveguides.
Changing the parameters of the elements allows this same circuit to
produce two nondegenerate down-converted photons that are entangled
in frequency or simultaneously entangled in frequency and
polarization. The second photonic circuit is designed to produce
modal entanglement by distinguishing the photons on the basis of
their frequencies. A modified version of this circuit can be used to
generate photons that are doubly entangled in mode number and
polarization. The third photonic circuit is designed to manage
dispersion by converting modal, spectral, and polarization
entanglement into path entanglement.
\end{abstract}

\section{Introduction}
Entanglement in a pair of photons can be manifested in degrees of
freedom that are discrete, such as polarization \cite{bennett84}, or
in degrees of freedom that are continuous, such as spectral or
spatial \cite{klyshko80,joobeur94,joobeur96,saleh00}. However, a
continuous degree of freedom may be binarized by selection of a pair
of values, as in path entanglement \cite{rarity90,franson99}, or, as
in orbital-angular-momentum entanglement \cite{mair01}, by expansion
in some basis while retaining only two basis functions
\cite{Walborn05}. Alternatively, in place of truncation, the space
of continuous functions of position may be mapped onto a binary set
of even and odd functions, as in spatial-parity entanglement
\cite{abouraddy07,yarnall07a,yarnall07b}. We recently investigated
the possibility of using spontaneous parametric down-conversion
(SPDC) in two-mode planar and circular waveguides to generate
guided-wave photon pairs entangled in mode number \cite{Saleh09}. If
the photons are confined in this manner, the spatial variables are
naturally binarized and can be used to represent a modal qubit
\cite{saleh10b}.

In this paper, we consider the design of photonic circuits, based on
Ti:LiNbO$_{3}$ diffused channel waveguides, that generate photon
pairs endowed with various combinations of modal, spectral, and
polarization entanglement. The on-chip generation of such photon
pairs can serve as a basic resource for one-way quantum computation
\cite{raussendorf01}, as well for linear optical quantum computing
\cite{knill01}. Down-converted photon pairs are generated via
spontaneous parametric down-conversion (SPDC) in a two-mode
waveguide using a CW pump source. The waveguide configuration
confines the photons to a single direction of propagation. Any of
the available degrees of freedom
--- mode number, frequency, or polarization --- can be used to
distinguish the down-converted photons while the others can serve as
carriers of entanglement. We study a class of photonic circuits that
comprises: 1) a nonlinear periodically poled two-mode waveguide
structure; 2) a set of single- and two-mode waveguide-based couplers
arranged in such a way that they suitably separate the three photons
comprising the SPDC process; and, for certain applications, 3) a
holographic Bragg grating.

A good deal of effort has been devoted to examining the properties
of single-mode silica-on-silicon waveguide quantum circuits
\cite{Politi08,Matthews09}, with an eye toward applications in
quantum-information processing
\cite{Bennett98,Nielsen00,OBrien09,Politi09,Cincotti09,ladd10}. For
these materials, the photon-generation process necessarily lies
off-chip, however. Single- and multi-mode potassium titanyl
phosphate (KTiOPO$_4$, KTP) waveguide structures have also been
extensively studied for producing pulsed spontaneous parametric
down-conversion
\cite{fiorentino07,Avenhaus09,Mosley09,zhong09,chen09}, but it
appears that only the generation process has been incorporated
on-chip. Earlier, periodically poled lithium niobate (PPLN)
waveguide structures were suggested for producing spontaneous
parametric down-conversion \cite{tanzilli01} and the conditions
required for generating counterpropagating entangled photons from an
unguided pump field were established \cite{booth02}. Furthermore,
the generation of non-collinear and non-degenerate
polarization-entangled photons via concurrent Type-I parametric
down-conversion was demonstrated in a PPLN crystal
\cite{dechatellus06}.

The use of lithium niobate photonic circuits has a number of merits:
1) the properties of the material are well-understood since it has
low loss and has long been the basis of integrated-optics technology
\cite{Nishihara89}, \cite[Chap.~8]{Saleh07}; 2) the material can
easily be periodically poled \cite{Busacca04,Lee04} for the purpose
of phase matching parametric interactions. And, as we show in this
paper: 3) circuit elements such as mode-separation components can
readily be designed for two-mode waveguides; and 4) the generation,
separation, and processing of entangled photons can all be
accommodated on a single chip. Moreover, consistency between
simulation and experimental measurement has been demonstrated in a
whole host of configurations
\cite{Alferness78,Alferness80,Hukriede03,Runde07,Runde08}. To
enhance tolerance to fabrication errors, photonic circuits can be
equipped with electro-optic adjustments. For example, an
electro-optically switched coupler with stepped phase-mismatch
reversal serves to maximize coupling between fabricated waveguides
\cite{schmidt76,kogelnik76}.

The paper is organized as follows. In Sec.~2, the theory of modal
entanglement in two-mode waveguides is reviewed. Section~3 is
dedicated to examining the operation of a photonic circuit that
produces a pair of frequency-degenerate photons at its output. As
explained in Sec.~4, changing the parameters of the elements
comprising the circuit considered in Sec.~3 allows it to produce
nondegenerate down-converted photons that are entangled in
frequency, or simultaneously entangled in frequency and
polarization. The second photonic circuit, examined in Sec.~5,
distinguishes the photons on the basis of their frequencies and
provides means for generating modal entanglement and, with some
modifications, modal and polarization double entanglement. The third
photonic circuit, considered in Sec.~6, is designed to manage the
deleterious effects of dispersion. The conclusion is provided in
Sec.~7.

\section{Generating modal entanglement in two-mode waveguides}

The theory underpinning the generation of modal entanglement in
waveguides has been recently described \cite{Saleh09}. Entanglement
arises from the multiple possibilities for satisfying energy and
momentum conservation, as required by the parametric interaction
process. In this section, we proceed to apply the theory provided in
\cite{Saleh09} to diffused channel Ti:LiNbO$_{3}$ waveguides.

Consider a SPDC process in a periodically poled two-mode waveguide
(TMW) structure, in which a cw pump wave $\,p\,$ propagating in the
$y$-direction is down-converted into a signal wave $\,s\,$ and an
idler wave $\,i$ \cite[Chap.~21]{Saleh07}. The fundamental mode of
this TMW ($m_{q}=0$) is even while the next-higher mode ($m_{q}=1$)
is odd, where $q=p,s,i$. We seek to generate an entangled state such
that if the signal is in the even mode, then the idler must be in
the odd mode, and \emph{vice versa}.

The biphoton state is written as \cite{Saleh09}
\begin{equation}
\vert \Psi\rangle\thicksim \displaystyle \int \:\mathrm{d}\omega_{s}\;\left[\Phi_{0,1,\sigma_{s},\sigma_{i}}\left(\omega_{s}\right)\:\vert \omega_{s},0,\sigma_{s}\rangle\vert\omega_{i},1,\sigma_{i} \rangle+ \,\Phi_{1,0,\sigma_{s},\sigma_{i}}\left( \omega_{s}\right)
 \:\vert\omega_{s},1, \sigma_{s}\rangle\vert\omega_{i},0,\sigma_{i}\rangle\right],
\label{eqEntangledstate}
\end{equation}
where the squared-magnitude of
$\Phi_{m_{s},m_{i},\sigma_{s},\sigma_{i}}$ represents the SPDC
spectrum  associated with the $ \left(
m_{s},m_{i},\sigma_{s},\sigma_{i}\right) $ component. The signal and
idler mode numbers are $m_{s}$ and $m_{i}$, their polarization
indexes are $\sigma_{s}$ and $\sigma_{i}$, and their angular
frequencies are $\omega_{s}$ and $\omega_{i}$, respectively, with $\omega_{i}=\omega_{p}-\omega_{s}$. The
angular frequency of the pump is denoted $\omega_{p}$. For ease of
display, the output spectra for the two possibilities represented by
the quantities $\left|\Phi_{0,1,
\sigma_{s},\sigma_{i}}\left(\omega_{s}\right)\right|^{2}$ and
$\left|\Phi_{1,0,
\sigma_{s},\sigma_{i}}\left(\omega_{s}\right)\right|^{2}$ are
normalized to the maximum of their peak values. Moreover, these
quantities are plotted only for signal frequencies above the
degenerate frequency $\frac12\omega_{p}$ by virtue of the fact that
$\Phi_{0,1,\sigma_{i},\sigma_{s}}\left(-\omega_{s}\right)=\Phi_{1,0,\sigma_{s},\sigma_{i}}\left(\omega_{s}\right)$.

This particular state requires that: 1) the pump mode be odd so that
the spatial overlap integral does not vanish; and 2) the phase
matching conditions of the two possibilities at the preselected
frequencies $ \overline{\omega}_{s} $ and $ \overline{\omega}_{i} $
be satisfied, i.e.,
\begin{equation}
\begin{array}{ll}
\Delta\beta_{0,1,\sigma_{s},\sigma_{i}}\left(\overline{\omega}_{s}\right)-2\pi k/\Lambda^{(k)}&\!\!\!\!=0\\
\Delta\beta_{1,0,\sigma_{s},\sigma_{i}}\left(\overline{\omega}_{s}\right)-2\pi
k/\Lambda^{(k)}&\!\!\!\!=0\,,
\end{array}
\label{eqPolingperiods}
\end{equation}
where $\Delta\beta_{m_{s},m_{i},\sigma_{s},\sigma_{i}}=
\beta_{m_{p},\sigma_{p}}\left(\omega_{p}\right)-\beta_{m_{s},
\sigma_{s}} \left(\overline{\omega}_{s}\right)
-\beta_{m_{i},\sigma_{i}}\left(\overline{\omega}_{i}\right) $;\,\,
$\overline{\omega}_{i} =\omega_{p}-\overline{\omega}_{s} $;\,\,
$\beta_{m_{q},\sigma_{q}}$ is the propagation constant of wave
$q$;\,\, and $\Lambda^{(k)}$ is the $k$th-order uniform poling
period. The efficiency of the interaction decreases by a factor $
1/k^{2} $ for the $k$th-order poling period with respect to the
first-order one \cite{Fejer92}. The choice of the TMW width $w_{1}$
determines the propagation constants of the interacting modes. The
values of the waveguide width $w_{1}$ and poling period
$\Lambda^{(k)}$ selected should ensure that these conditions are met
or approximately met.

The indistinguishability between the down-converted photons is
typically ascertained via a Hong--Ou--Mandel (HOM) interferometer
\cite{Hong87,Campos90}. The rate $R(\tau)$ of photon coincidences at
a pair of detectors placed at the two output ports of the
interferometer is given by~\cite{Saleh09}
\begin{equation}
R\left(\tau \right)=
\displaystyle\int_{-\infty}^{\infty}\mathrm{d}\omega_{s}\;
\biggl\lbrace\left\vert\Phi_{0,1,\sigma_{s},\sigma_{i}}\left(
\omega_{s}\right)
\right\vert^{2}-\Phi^{}_{0,1,\sigma_{s},\sigma_{i}}\left(
\omega_{s}\right)\, \Phi_{0,1,\sigma_{s},\sigma_{i}}^{*}\left(
-\omega_{s}\right)\:\exp\left[j\left(\omega_{p}-
2\omega_{s}\right)\tau \right]\bigg\rbrace , \label{coincidence}
\end{equation}
where  $\tau$ is the temporal delay between the down-converted
photons and the asterisk denotes complex conjugation.

\begin{figure}[t]
\centering
\includegraphics[width=2in,totalheight =1.2in]{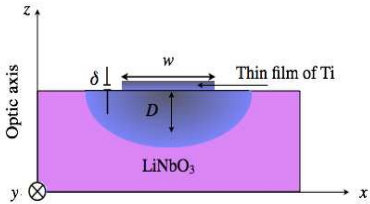}
\caption{Cross-sectional view of the fabrication of a diffused
channel Ti:LiNbO$_{3}$ waveguide (not to scale). A thin film of
titanium of thickness $\delta = 100$ nm and width $w$ is diffused
into a $z$-cut, $y$-propagating LiNbO$_3$ crystal. The diffusion
length $D = 3\: \mu$m.} \label{ChannelWGs}
\end{figure}

All of the simulations presented in this paper refer to structures
that make use of Ti:LiNbO$_{3}$ diffused channel waveguides, as
illustrated in Fig.~\ref{ChannelWGs}. These waveguides are
fabricated by diffusing a thin film of titanium (Ti), with thickness
$ \delta = 100$ nm and width $w$, into a $z$-cut, $y$-propagating
LiNbO$_{3}$ crystal. The diffusion length $D$ is taken to be the
same in the two transverse directions: $D = 3\: \mu $m. The TE mode
polarized in the $x$-direction sees the ordinary refractive index
$n_{o}$, whereas the TM mode polarized in the $z$-direction sees the
extraordinary refractive index $n_{e}$.

The ordinary and extraordinary refractive indexes may be calculated
by making use of the Sellmeier equations \cite[Chap.~5]{Saleh07},
\cite{jundt97,Wong02}. The refractive-index increase introduced by
titanium indiffusion is characterized by $\Delta
n=2\delta\rho\,\,\mathrm{erf}\!\left( w/2D\right) /\sqrt{\pi}\,D $,
where $ \rho = 0.47$ and $0.625$ for $n_{o}$ and $n_{e}$,
respectively \cite{Feit83}. To accommodate wavelength dispersion,
$\Delta n$ can be modified by incorporating the weak factor $\xi =
0.052 + 0.065/\lambda^{2}$, where the wavelength $\lambda$ is
specified in $\mu$m \cite{Hutcheson87}. We calculate the effective
refractive index $n_{\mathrm{eff}}$ of a confined mode in two ways:
1) by using the effective-index method described in \cite{Hocker77};
and 2) by making use of the commercial photonic and network design
software package RSoft. The propagation constant of a guided mode is
related to $n_{\mathrm{eff}}$ via $\beta = 2\pi
n_{\mathrm{eff}}/\lambda$.

In the following sections, we consider several photonic circuits for
the generation, separation, and control of down-converted photons.
In Sec.~3, we consider a circuit that generates degenerate photons
in separated single-mode waveguides. In this case, the two
conditions specified in (\ref{eqPolingperiods}) collapse to a single
condition since $\overline{\omega}_{s} =\overline{\omega}_{i}$ and
$\sigma_{s}=\sigma_{i}$. In Sec.~4,
with a change of parameters, this same circuit is used to produce
nondegenerate photons entangled in frequency or simultaneously
entangled in frequency and polarization. Section~5 relates to
photonic circuitry that produces nondegenerate photons distinguished
by their frequencies, which leads to modal as well as modal and
polarization double entanglement.

\section{Photonic circuit for generating degenerate photons in different modes}
This section is dedicated to examining the operation of a photonic
circuit that produces a pair of down-converted photons, and then
uses their differing spatial properties (mode numbers) to ultimately
guide them into two separate, single-mode waveguides. The output
provides a pair of degenerate photons that have even parity and can
be fed into an HOM interferometer \cite{Hong87,Campos90} integrated
on the same chip (see Sec.~6).

\begin{figure}[t]
\centering
\includegraphics[width=3.4 in,totalheight =3.4 in]{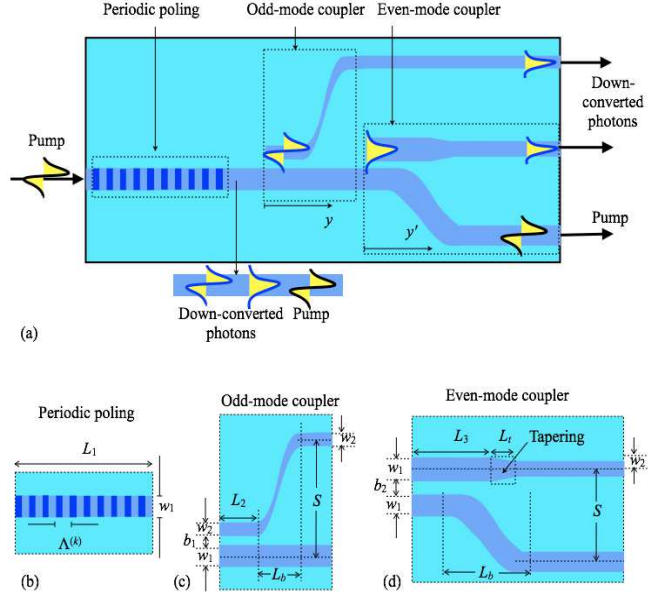}
\caption{(a) Sketch of a photonic circuit that can be used to
generate two photons in different single-mode waveguides (not to
scale). The circuit comprises three sequential stages: a
periodically poled region, an odd-mode coupler, and an even-mode
coupler. The parameters associated with these stages are defined in
panels (b), (c), and (d), respectively. (b) The periodically poled
waveguide has width $w_{1}$, length $L_{1}$, and a $k$th-order
poling period $\Lambda^{(k)}$. (c) The odd-mode coupler is
implemented by bringing a single-mode waveguide (SMW) of width
$w_{2}$ and length $L_{2}$ into proximity with a two-mode waveguide
(TMW) of width $w_{1}$. The two waveguides are separated by a
distance $b_1$. An $S$-bend waveguide of initial and final width
$w_{2}$, and bending length $L_{b}$, is attached to the end of the
SMW. The center-to-center separation between the output of the
$S$-bend and the TMW is denoted $S$. All $S$-bends considered in
this paper have dimensions $L_{b} = 10$ mm and $S = 127 \:\mu$m (the
standard spatial separation \cite{Runde08}). (d) The even-mode
coupler is implemented by bringing a second TMW, of width $w_{1}$
and length $L_{3}$, into proximity with the original TMW. The two
waveguides are separated by a distance $b_2$. The end of this second
TMW is adiabatically tapered over a length $L_{t}$ so that it
matches the width $w_{2}$ at the output of the odd-mode coupler. The
length of the taper $L_{t} = 1.5$ mm, as it is throughout this
paper.} \label{CircuitI}
\end{figure}

In particular, we consider the implementation of degenerate Type-0
$(e,e,e)$ SPDC in a TMW of width $w_{1}$, wherein a pump photon with
$m_{p}=1$ is split into a pair of down-converted photons with
different spatial parities. The associated two-photon state is
\begin{equation}
\vert \Psi\rangle\thicksim \displaystyle \int \:\mathrm{d}\omega_{s}\;\left[\Phi_{0,1,e,e}\left(\omega_{s}\right)
\:\vert \omega_{s},e\rangle_{0}\,\vert\omega_{p}-\omega_{s},e \rangle_{1}+\,\Phi_{1,0,e,e}\left( \omega_{s}\right)
 \:\vert\omega_{s},e\rangle_{1}\,\vert\omega_{p}-\omega_{s},e\rangle_{0}\right],
\end{equation}
where the two terms merge for the degenerate case, i.e., when
$\omega_{s}=\frac12\omega_{p}$. Figure~\ref{CircuitI}(a) represents
a photonic circuit that can be used to generate photon pairs of this
type. The circuit comprises three sequential stages: a periodically
poled region, an odd-mode coupler, and an even-mode coupler. We
consider these three stages in turn.

The \textit{periodically poled stage} shown in
Fig.~\ref{CircuitI}(b) is characterized by its width $w_{1}$, length
$L_{1}$, and $k$th-order poling period $\Lambda^{(k)}$. For certain
values of the width, there will always be a single value of the
poling period that satisfies (\ref{eqPolingperiods}) so that there
is flexibility in choosing the TMW width $w_{1}$. Once the width is
assigned, the poling period is determined via
(\ref{eqPolingperiods}). There are no restrictions on the length of
the nonlinear stage $L_{1}$, save those associated with the process
of fabrication. Increasing $L_{1}$ does, of course, increase the
flux of photon pairs but we will see in the next Section that there
is a limitation on how large $L_{1}$ can be made for generating
entangled photons.

The \textit{odd-mode coupler} is used to extract only the
down-converted photon with an odd spatial distribution. Its
principle of operation is based on selective coupling between
adjacent waveguides of different widths. The even and odd modes of
the TMW are characterized by different propagation constants. As
depicted in Fig.~\ref{CircuitI}(c), an auxiliary single-mode
waveguide (SMW) with appropriate width $w_{2}$, length $L_{2}$, and
separation distance $b_{1}$ from the TMW can be used to phase-match
the odd-mode in the TMW to the even-mode in the SMW. Attached to the
end of the SMW is an $S$-bend waveguide in which the initial and
final widths are both $w_{2}$; this obviates the possibility of
further unwanted coupling to the TMW. In short, the odd-mode coupler
distinguishes between the two down-converted photons based on their
spatial profiles and delivers the initially odd-mode photon as an
even-mode photon at its output.

The \textit{even-mode coupler} is used to separate the
down-converted photon with an even spatial profile from the pump
wave. The principle of operation is based on the limited extent of
the short-wavelength evanescent pump field, which precludes it from
coupling to other waveguides of the same width located at typical
distances. The result is that the even-mode down-converted photon
alone is coupled into a second identical TMW, of length $L_{3}$ and
separation distance $b_{2}$ from the original TMW, as shown in
Fig.~\ref{CircuitI}(d). The end of the second TMW is then coupled to
a SMW of width $w_{2}$ via an adiabatically tapered waveguide region
of length $L_{t}$. The tapering offers two benefits: 1) the even
mode of the TMW is transformed into an even mode that matches that
of the partner photon at the output of the odd-mode coupler; and 2)
unwanted odd modes inadvertently coupled in from external radiation
are eliminated. The down-converted degenerate photons produced by
this photonic circuit have the same mode number at the output, and
can therefore interfere quantum mechanically \cite{Campos90}. It is
worthy of mention that heralded single-photon pure states in
well-defined spatiotemporal modes are required for many quantum
information technology applications \cite{saleh10b}, such as quantum
cryptography \cite{Gisin02} and linear optical quantum computing
\cite{knill01}.

\begin{figure}[t]
\centering
\includegraphics[width=3.4 in,totalheight =3.4 in]{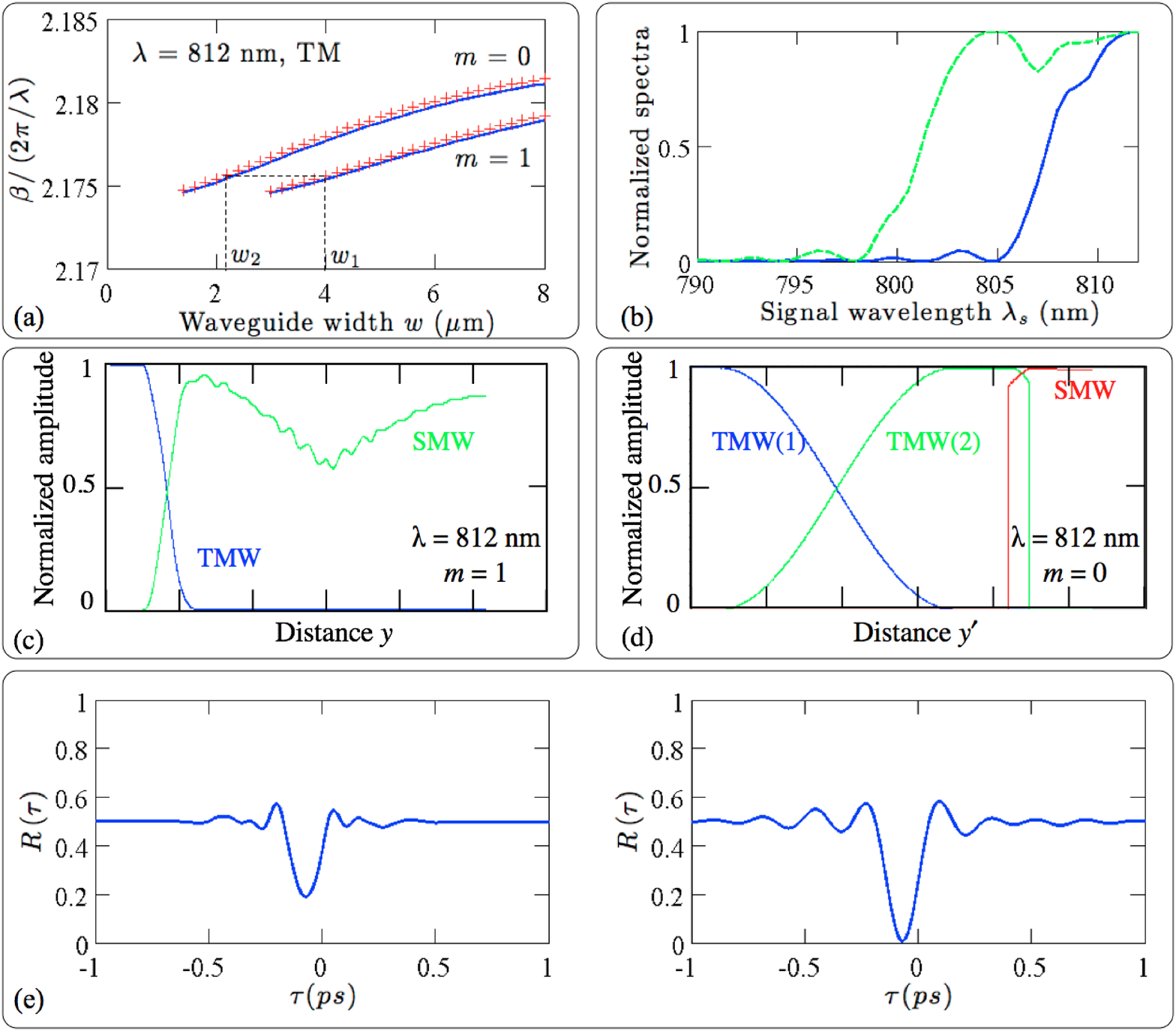}
\caption{Simulated performance of the photonic circuit presented in
Fig.~\ref{CircuitI}. A pump at wavelength $\lambda_{p}= 406$ nm, and
in an odd spatial mode ($m_{p}=1$), is used to produce degenerate
Type-0 $(e,e,e)$ down-conversion with an effective nonlinear optical
coefficient $d_{\mathrm{eff}}=d_{33}$ \cite{Myers95}. The circuit
dimensions are $w_{1}= 4 \: \mu$m, $w_{2}= 2.2 \: \mu$m, $L_{1}=20$
mm, $L_{2}=0.85$ mm, $L_{3}=21$ mm, and $b_{1} = b_{2}= 5 \:\mu$m.
The poling period $\Lambda^{(1)} =2.644 \: \mu $m is designed for
operation at a temperature of 80$^\circ$ C and can be temperature
tuned. Values for $L_b$, $L_t$, and $S$ are specified in the caption
of Fig.~\ref{CircuitI}. (a) Dependencies of the propagation
constants $\beta$ of the two modes for the down-converted photons on
the waveguide widths $w$. The solid curves were obtained using the
effective-index method described in \cite{Hocker77}, whereas the
plus signs were computed using the software package RSoft. The
dotted vertical lines represent the widths $w_1$ and $w_2$ that were
chosen. (b) Dependence of the normalized output spectra on the
wavelength of the signal $\lambda_{s}$. The solid blue and dotted
green curves represent $\left|\Phi_{0,1,e,e}\right|^{2}$ and
$\left|\Phi_{1,0,e,e}\right|^{2}$, respectively, for wavelengths
below the degenerate value of 812 nm. (c) Performance of the
odd-mode coupler. The blue and green curves represent the evolution
with distance of the normalized amplitudes of the odd mode in the
TMW and the even mode in the SMW, respectively. (d) Performance of
the even-mode coupler. The blue, green, and red curves represent the
evolution with distance of the normalized amplitudes of the even
modes in the original two-mode waveguide TMW(1), in the second
two-mode waveguide TMW(2), and in the SMW, respectively. The even
mode of TMW(2) is defined as extending to the end of the tapered
region, whereas that for the SMW is defined as initiating at the
beginning of the tapered region. (e) Dependencies of the normalized
coincidence rate $R$ on the temporal delay $\tau$, without insertion
(left curve), and with insertion (right curve), of two 10-nm
narrowband filters centered about 812 nm, before the two output
detectors.} \label{ExCorrelation}
\end{figure}
Figure~\ref{ExCorrelation} provides an example in which the
performance of this photonic circuit is simulated.
Figure~\ref{ExCorrelation}(a) displays the dependencies of the
fundamental and first-order mode propagation constants $\beta$ on
the waveguide widths $w$. These curves are used to select the widths
of the TMW and SMW; the horizontal dotted line represents the loci
of the phase-matching conditions between even and odd modes in
waveguides of different widths. Figure~\ref{ExCorrelation}(b) shows
the normalized output spectra $\left|\Phi_{0,1,e,e}\right|^{2}$  and
$\left|\Phi_{1,0,e,e}\right|^{2}$ as a function of the signal
wavelength $\lambda_{s}$. As a consequence of the large phase
mismatch between the interacting modes, the value for the
first-order poling period turns out to be small, but values in this
range have been realized using the method of surface poling in
diffused channel waveguides \cite{Busacca04}. Using a pump source of
longer wavelength, such as 532 nm, would permit regular poling
techniques to be used. A salutary feature is that the Type-0
interaction makes use of the strongest nonlinear component of the
second-order tensor, $d_{33}$, so that high efficiency is expected
from this interaction \cite{Myers95}. Figures~\ref{ExCorrelation}(c)
and \ref{ExCorrelation}(d) display the evolution with distance of
the normalized amplitudes of the interacting modes in the odd- and
even-mode couplers, respectively. For the odd-mode coupler, the
amplitude of the even excited mode in the SMW exhibits a dip that is
associated with the tapered nature of the $S$-bend. The values $m =
1$ and 0 in Figs.~\ref{ExCorrelation}(c) and \ref{ExCorrelation}(d),
respectively, refer to the mode numbers at the point of generation.
Figure~\ref{ExCorrelation}(e) displays the dependencies of the
normalized coincidence rate $R$ on the temporal delay $\tau$,
without insertion (left curve), and with insertion (right curve), of
two narrowband filters at the degenerate frequency before the two
output detectors. The reasons underlying the reduced visibility of
the dip, and its displacement from $\tau=0$, are considered in
Sec.~V of \cite{Saleh09}.

\section{Photonic circuits for generating spectral and spectral / polarization entanglement}
Changing the values of the parameters provided in
Fig.~\ref{ExCorrelation} allows the circuit shown in
Fig.~\ref{CircuitI} to produce two nondegenerate down-converted
photons that are entangled in frequency, or simultaneously in
frequency and polarization, via a Type-0 or Type-II interaction,
respectively.

Using the mode numbers to distinguish the photons,
(\ref{eqEntangledstate}) takes the form
\begin{equation}
\vert \Psi\rangle\thicksim \displaystyle \int \:
\mathrm{d}\omega_{s}\; \left[\Phi_{0,1,\sigma_{s},\sigma_{i}}\left(
\omega_{s}\right)
\:\vert \omega_{s},\sigma_{s}\rangle_{0}\vert\omega_{i},\sigma_{i} \rangle_{1}+\:\Phi_{1,0,\sigma_{s},\sigma_{i}}\left( \omega_{s}\right)
 \:\vert\omega_{s},
 \sigma_{s}\rangle_{1}\vert\omega_{i},\sigma_{i}\rangle_{0}\right].
\end{equation}
However, some restrictions apply to the generation and separation
portions of the circuit, as discussed in the following paragraphs.

With respect to the generation region, values of the waveguide width
$w$ and poling period $\Lambda^{(k)}$ govern how the conditions in
(\ref{eqPolingperiods}) are satisfied. Simulations show that: 1) for
certain values of the waveguide width, slightly different poling
periods are required to satisfy these conditions; and 2) the
absolute difference between the quantities $ \left[
\beta_{0,\sigma_{s}}\left(\overline{\omega}_{s}\right)
+\beta_{1,\sigma_{i}}\left(\overline{\omega}_{i}\right)\right]$ and
$\left[ \beta_{0,\sigma_{i}}\left(\overline{\omega}_{i}\right)
+\beta_{1,\sigma_{s}}\left(\overline{\omega}_{s}\right) \right]$,
which appear in (\ref{eqPolingperiods}), is \emph{nearly}
independent of waveguide width. Consequently, the choice of $w_{1}$
can be left to other considerations, such as fabrication limitations
and subsequent design requirements.

Our strategy is to satisfy the sum condition, rather than each
condition individually, for a specified value of the waveguide
width, so that
\begin{equation}
\Delta\beta_{\rm avg}\left(\overline{\omega}_{s}\right)-2\pi
k/\Lambda^{(k)}=0,
\end{equation}
where $\Delta\beta_{\rm
avg}=\frac12\left[\Delta\beta_{0,1,\sigma_{s},\sigma_{i}}\left(\overline{\omega}_{s}\right)
+\Delta\beta_{1,0,\sigma_{s},\sigma_{i}}\left(\overline{\omega}_{s}\right)
\right]$. Substituting the value of the poling period
$\Lambda^{(k)}$ obtained in both conditions of
(\ref{eqPolingperiods}) leads to small errors in satisfying the
phase-matching conditions for the two interactions. These errors can
be used to determine the coherence lengths for the two interactions,
which then provide upper bounds to the length of the periodically
poled region $L_{1}$. The choice of $L_{1}$ must also result in good
spectral overlap between $\left|\Phi_{0,1,
\sigma_{s},\sigma_{i}}\left(\omega_{s}\right) \right|^{2}$ and
$\left|\Phi_{1,0,
\sigma_{s},\sigma_{i}}\left(\omega_{s}\right)\right|^{2}$ around the
preselected frequencies, since $ L_{1} $ determines the widths of
these functions. It turns out that this strategy leads to almost
maximally entangled states at the preselected frequencies.

\begin{figure}[t]
\centering
\includegraphics[width=3.4 in,totalheight =5 in]{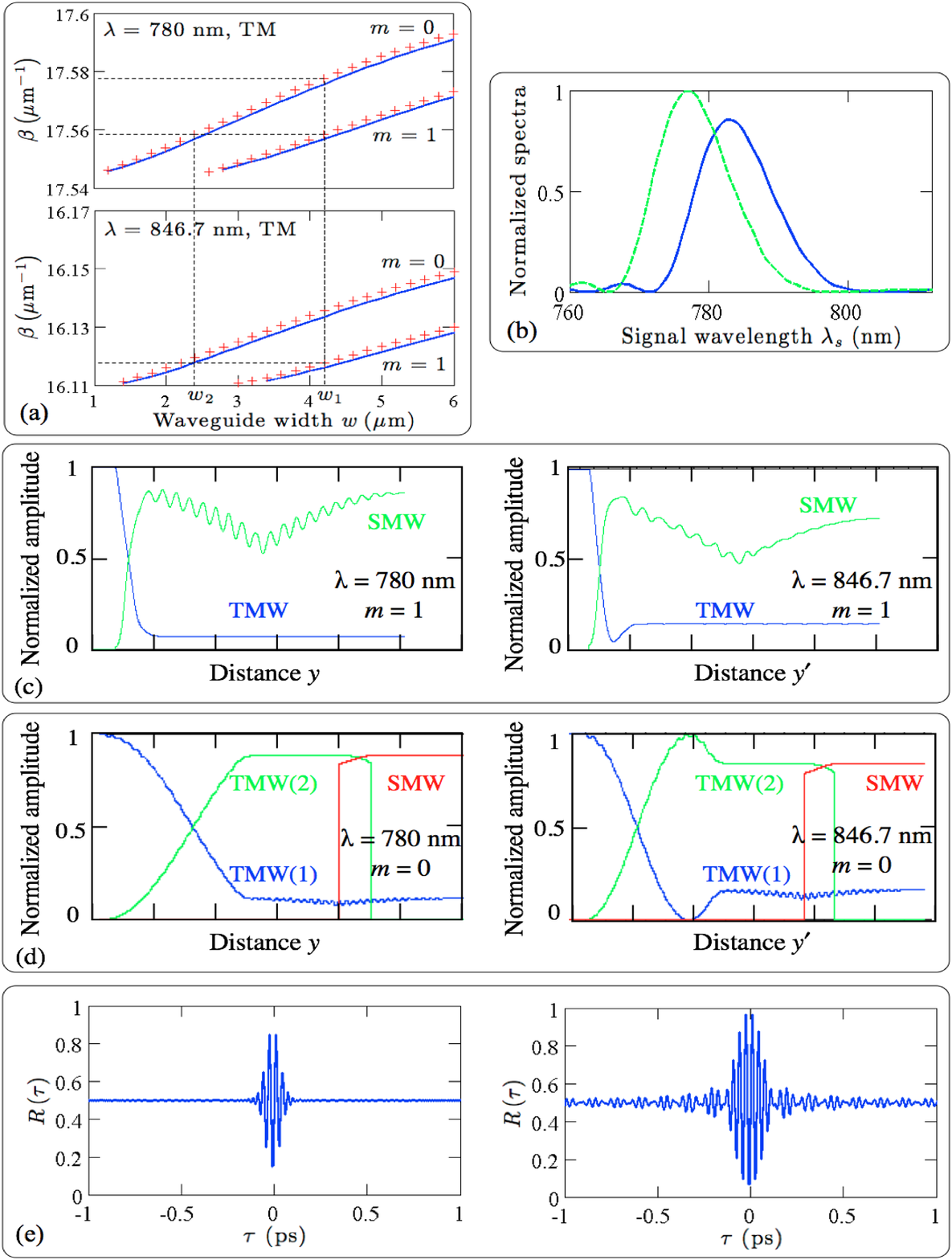}
\caption{Simulated performance of the photonic circuit portrayed in
Fig.~\ref{CircuitI} when configured to generate spectrally entangled
photons. An odd-mode ($m_{p}=1$) TM laser pump at $\lambda_{p}=406$
nm is used to generate nondegenerate Type-0 $(e,e,e)$ SPDC with a
nonlinear optical coefficient $d_{\mathrm{eff}}=d_{33}$. The center
wavelengths of the two down-converted photons are 780 nm and 846.7
nm. The circuit dimensions are $ w_{1}= 4.2 \;\mu$m, $w_{2}= 2.4 \:
\mu$m, $L_{1}=2$ mm, $L_{2}=0.18$ mm, $L_{3}=11$ mm, and $b_{1}=
b_{2}= 4 \:\mu$m. The poling period, $\Lambda^{(1)} = 2.588 \:
\mu$m, is designed for operation at a temperature of 80$^\circ$ C
and can be temperature tuned. The panels in this figure are similar
to those displayed in Fig.~\ref{ExCorrelation}, except that panels
(a), (c), and (d) each contain separate plots for the two
wavelengths. Furthermore, in panel (e), the two filters placed in
front of the detectors are centered about 780 nm and 846.7 nm,
instead of both being centered about 812 nm, as for degenerate
down-conversion. } \label{ExSpectral}
\end{figure}
With respect to the coupling region, the design must optimize
coupling at both preselected frequencies, $\overline{\omega}_{s}$
and $\overline{\omega}_{i}$. The choice of $w_{1}$ and $w_{2}$ for
the odd-mode coupler is based on satisfying the phase matching
conditions between an odd mode in a TMW of width $w_{1}$ and an even
mode in a SMW of width $w_{2}$, for both $\overline{\omega}_{s}$ and
$\overline{\omega}_{i}$:
\begin{equation}
\begin{array}{c}
 \beta_{1,\sigma_{s}}\left(\overline{\omega}_{s},\,w_{1}\right)=\beta_{0,\sigma_{s}}\left(\overline{\omega}_{s},\,w_{2}\right),  \\
 \beta_{1,\sigma_{i}}\left(\overline{\omega}_{i},\,w_{1}\right)=\beta_{0,\sigma_{i}}\left(\overline{\omega}_{i},\,w_{2}\right) .
\end{array}
\label{eqOddCoupler}
\end{equation}
Since the down-conversion generation process depends only weakly on
waveguide width, we select the TMW width in the generation region to
be equal to $w_{1}$; this obviates the need to taper the waveguide
from the periodically poled region to the odd-mode coupler. The
coupling length is frequency and polarization dependent, however, so
that the SMW length $L_{2}$, in combination with the separation
distance $b_{1}$, must be carefully chosen to maximize the coupling
for both $\overline{\omega}_{s}$ and $\overline{\omega}_{i}$. The
same considerations apply for the even-mode coupler, so that the
second TMW length $L_{3}$ and separation distance $ b_{2}$ are
selected similarly.

\begin{figure}[t]
\centering
\includegraphics[width=3.4 in,totalheight =5 in]{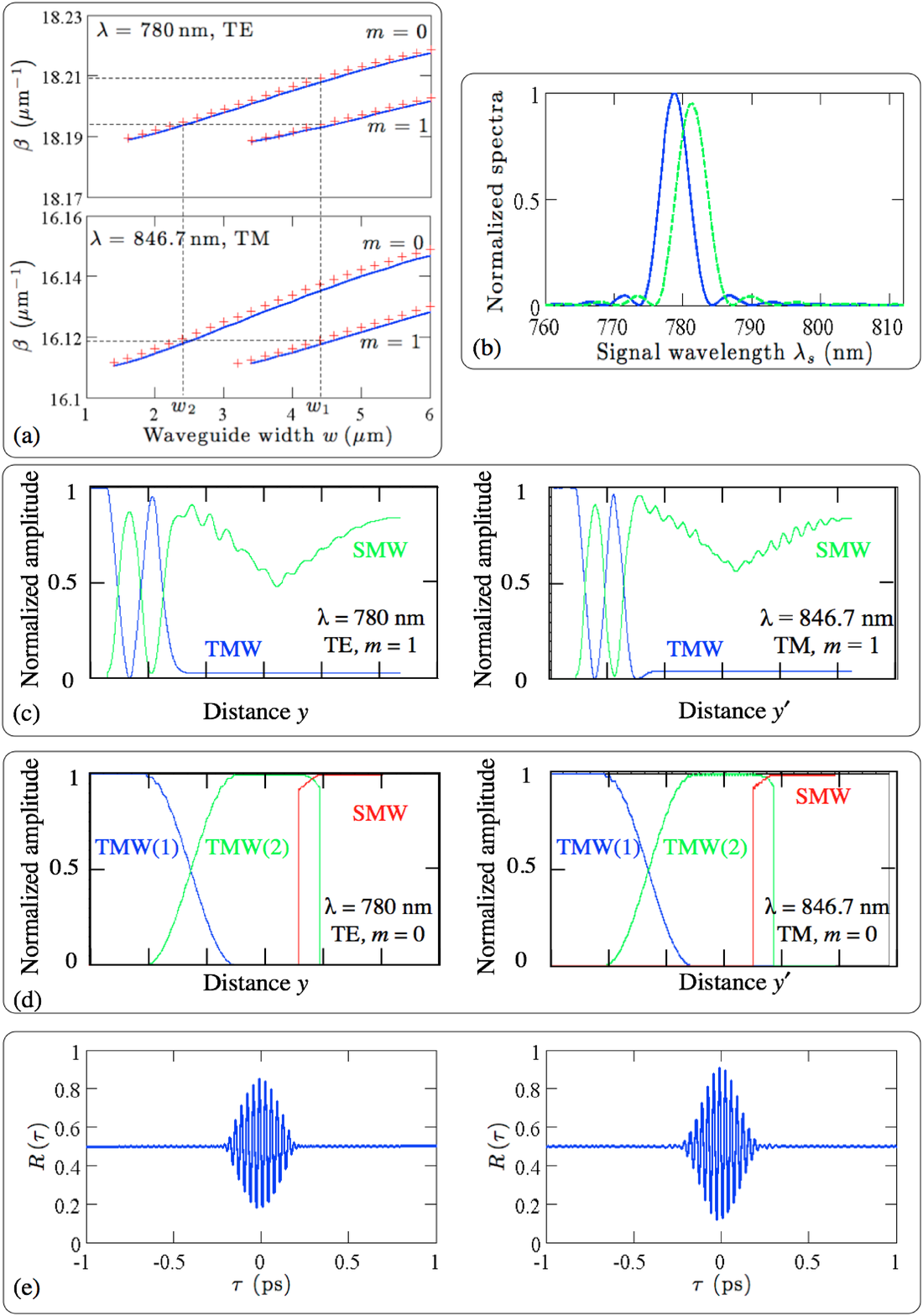}
\caption{Simulated performance of the photonic circuit displayed in
Fig.~\ref{CircuitI} when configured to generate photons
simultaneously entangled in frequency and polarization. A laser pump
at $\lambda_{p}=406$ nm, in mode $m_{p}=1$, is used to produce
nondegenerate Type-II $(o,e,o)$ SPDC, with a nonlinear optical
coefficient $d_{\mathrm{eff}}=d_{31}\ll d_{33}$ \cite{Myers95}. A TE
pump is used in place of the TM pump employed in
Fig.~\ref{ExSpectral}, and $\Lambda^{(k)}$ is chosen appropriately.
The center wavelengths of the two down-converted photons are 780 nm
and 846.7 nm. The circuit dimensions are $ w_{1}= 4.4 \: \mu$m,
$w_{2}= 2.4 \: \mu$m, $ L_{1}=1 $ mm, $ L_{2}=1.95 $ mm, $L_{3}=11$
mm, and $ b_{1}= b_{2}= 4 \:\mu$m. The poling period, $
\Lambda^{(1)} =1.869 \:\mu$m, is designed for operation at a
temperature of 80$^\circ$ C and can be temperature tuned. All panels
are similar to those displayed in Fig.~\ref{ExSpectral} for
nondegenerate Type-0 $(e,e,e)$ SPDC.} \label{ExPolarization}
\end{figure}
An example, provided in Fig.~\ref{ExSpectral}, illustrates the
simulated performance of the photonic circuit shown in
Fig.~\ref{CircuitI} for generating spectrally entangled photons.
Consider nondegenerate Type-0 $(e,e,e)$ SPDC, in which a pump laser
of wavelength 406 nm, in mode $m_{p}=1$, generates two photons
centered about the wavelengths 780 nm and 846.7 nm. The TMW width
$w_{1}$ and the SMW width $w_{2}$ are determined via
Fig.~\ref{ExSpectral}(a), which represents the dependencies of the
propagation constants on waveguide width. The values $w_{1}$ and
$w_{2}$ are chosen to satisfy (\ref{eqOddCoupler}) as closely as
possible. Figure~\ref{ExSpectral}(b) displays the output normalized
spectra, $\left|\Phi_{0,1, e,e}\left(\omega_{s}\right)\right|^{2}$
and $\left|\Phi_{1,0, e,e} \left(\omega_{s}\right)\right|^{2}$, for
an SPDC poling length $L_{1}$ that was selected to attain maximum
entanglement and high efficiency. The performance of the odd- and
even-mode couplers are displayed in Figs.~\ref{ExSpectral}(c) and
\ref{ExSpectral}(d), respectively; each panel contains plots for the
two frequencies $\overline{\omega}_{s}$ and $\overline{\omega}_{i}$.
The values $m = 1$ and 0 in Figs.~\ref{ExSpectral}(c) and
\ref{ExSpectral}(d), respectively, refer to the mode numbers at the
point of generation, and the curves are color-coded in the same way
as the corresponding curves in Fig.~\ref{ExCorrelation}. The values
of $L_{2}$ ($L_{3}$) and $b_{1}$ ($b_{2}$) are selected to optimize
coupling of the odd (even) modes for both $\overline{\omega}_{s}$
and $\overline{\omega}_{i}$. Figure~\ref{ExSpectral}(e) displays the
behavior of the normalized coincidence rate function $R(\tau)$ for
the down-converted photons, without insertion (left curve), and with
insertion (right curve), of two 10-nm narrowband filters, centered
about 780 and 846.7 nm, before the two output detectors.

An example based on nondegenerate Type-II $(o,e,o)$ SPDC,
illustrated in Fig.~\ref{ExPolarization}, suggests that simultaneous
spectral and polarization entanglement can be obtained by again
making use of the photonic circuit shown in Fig.~\ref{CircuitI}. By
convention, the notation $(\cdot\,,\cdot\,,\cdot)$ indicates, in
consecutive order, the polarization of the down-converted photon
whose frequency lies above the degenerate frequency, the
down-converted photon whose frequency lies below the degenerate
frequency, and the pump photon. Type-II behavior is obtained by
using a TE pump in place of the TM pump employed in
Fig.~\ref{ExSpectral}, and by choosing the circuit-element
parameters, including $\Lambda^{(k)}$, appropriately. The panels
displayed in Fig.~\ref{ExPolarization} are similar to those shown in
Fig.~\ref{ExSpectral}. Again, the values $m = 1$ and 0 in
Figs.~\ref{ExPolarization}(c) and \ref{ExPolarization}(d),
respectively, refer to the mode numbers at the point of generation,
and the curves are color-coded in the same way as the corresponding
curves in Figs.~\ref{ExCorrelation} and \ref{ExSpectral}. In
principle, spectral and polarization double entanglement
\cite{Saleh09} could also be attained with this arrangement, but the
parameters that achieve it are not easily established.

\section{Photonic circuits for generating modal and double modal / polarization entanglement}

This section focuses on the operation of a photonic circuit that
generates modal entanglement by distinguishing two nondegenerate
photons on the basis of their frequencies, rather than their mode
numbers. Modally entangled photons in a two-mode waveguide have the
following property: a photon whose frequency lies above the
degenerate frequency and appears in the fundamental (even) mode is
always accompanied by another whose frequency lies below the
degenerate frequency and appears in the first-order (odd) mode, and
\emph{vice versa}. Alternatively, the photons could be distinguished
on the basis of their polarizations in the context of a Type-II
configuration. When the photons are distinguished on the basis of
their frequencies, the biphoton state (\ref{eqEntangledstate}) is
written as
\begin{equation}
\vert \Psi\rangle\thicksim \displaystyle \int \:
\mathrm{d}\omega_{s}
\;\left[\Phi_{0,1,\sigma_{s},\sigma_{i}}\left( \omega_{s}\right)
\:\vert  0,\sigma_{s}\rangle_{\omega_{s}}\vert 1,\sigma_{i} \rangle_{\omega_{i}}+\:\Phi_{1,0,\sigma_{s},\sigma_{i}}\left( \omega_{s}\right)
 \:\vert 1, \sigma_{s}\rangle_{\omega_{s}}\vert0,\sigma_{i}\rangle_{\omega_{i}}\right].
\end{equation}

\begin{figure}[t]
\centering
\includegraphics[width=3.4 in,totalheight =3.2in]{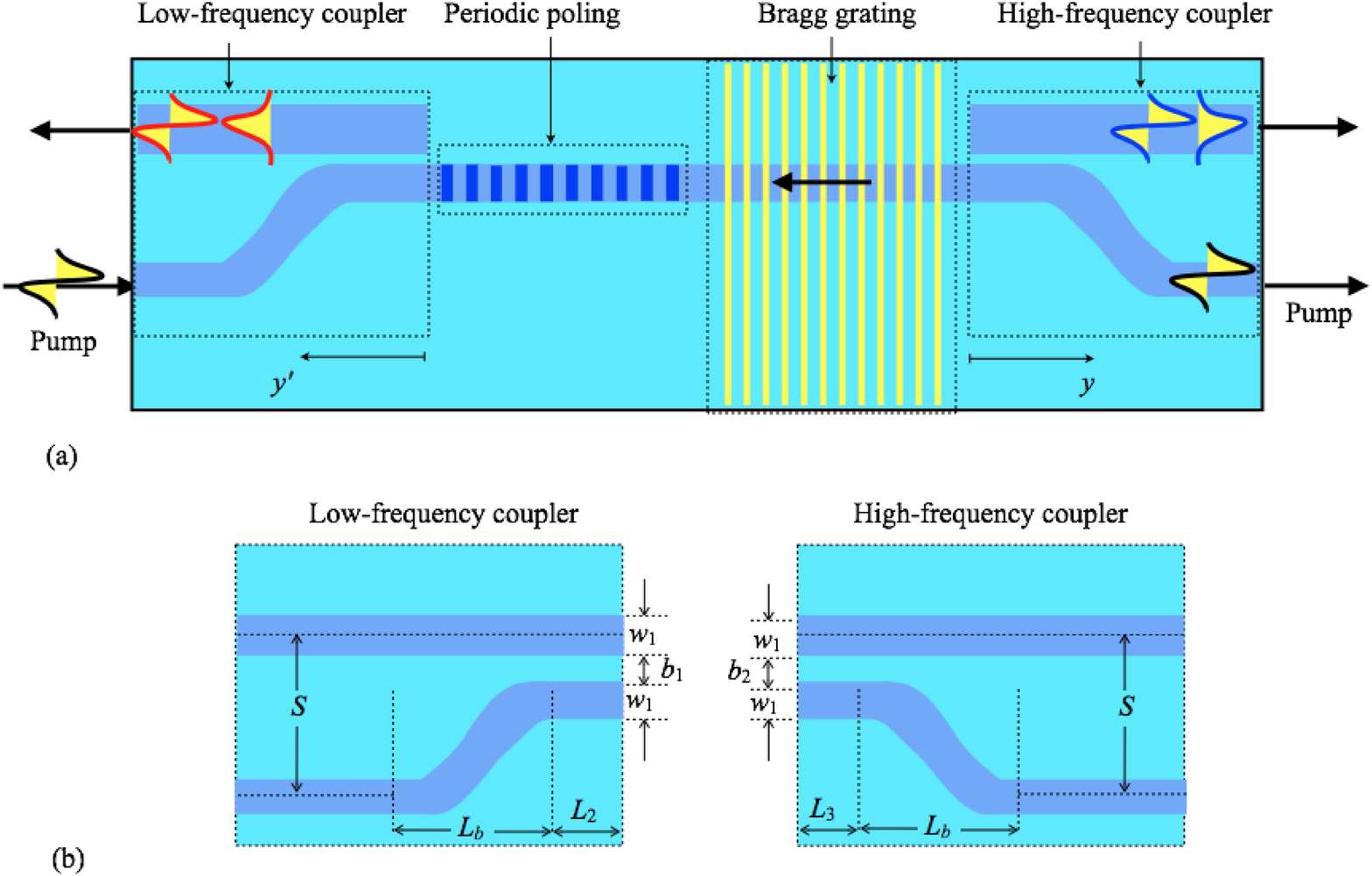}
\caption{(a) Sketch of a photonic circuit that can be used to
generate two photons entangled in mode number (not to scale). The
circuit comprises three stages: a periodically poled region, a
holographic Bragg grating, and two couplers (a low-frequency coupler
located to the left and a high-frequency coupler located to the the
right). The periodically poled waveguide is identical to that
portrayed in Fig.~\ref{CircuitI}(b), with width $w_{1}$, length
$L_{1}$, and poling period $\Lambda^{(k)}$. The holographic Bragg
grating reflects the low-frequency photon and transmits the
high-frequency photon. (b) Definitions of the parameters associated
the couplers. As in Fig.~\ref{CircuitI}(c), $L_{b} = 10$ mm and $S =
127 \:\mu$m.} \label{CircuitII}
\end{figure}
Modal entanglement is generated by the photonic circuit displayed in
Fig.~\ref{CircuitII}. It comprises three stages. The first
\emph{periodically poled stage} is a nonlinear region, as in
Fig.~\ref{CircuitI}(b), that produces a pair of nondegenerate
down-converted photons via SPDC. As discussed in Sec.~3, this stage
is characterized by its width $w_{1}$, length $L_{1}$, and
$k$th-order poling period $\Lambda^{(k)}$. The choice of these
parameters is governed by the same considerations as those attendant
to the discussion in Sec.~4.

The second stage is a \emph{holographic Bragg grating} that
distinguishes the down-converted photons based on their frequency,
such as by reflecting the low-frequency photon while transmitting
the high-frequency photon so they travel in opposite directions. The
third stage consists of two couplers that transfer the even and odd
down-converted photons from the original waveguide in which they are
generated to the output-port waveguides of the circuit, via a
\emph{low-frequency coupler} to the left and a \emph{high-frequency
coupler} to the right.

The holographic Bragg grating can be inscribed on the Ti:LiNbO$_{3}$
waveguide by superposing two coherent plane waves and making use of
proton exchange \cite{Ferriere06}, or a dopant such as Cu together
with thermal fixing \cite{Hukriede03}. The sinusoidal
light-intensity pattern generated by the superposition of the plane
waves \cite[Chap.~2]{Saleh07} is impressed on the waveguide as a
sinusoidal modulation of the refractive index. The wavelength of
light at which the reflected intensity is maximum (for normal
incidence) is $\lambda=2n_{\mathrm{eff}}\,\Lambda_{B}$, where
$\Lambda_{B}$ is the grating period and $n_{\mathrm{eff}}$ is the
effective refractive index of the guided mode. The bandwidth of the
reflected light decreases with increasing grating length. The
fundamental and first-order modes can be reflected at a given
wavelength by making use of either a single hologram of short length
or two mutliplexed holograms of longer length. Multiplexing is most
effectively achieved by recording the holograms sequentially
\cite{Hukriede03}.

The low- and high-frequency couplers are designed to couple the
idler-frequency ($\overline{\omega}_{i}$) and signal-frequency
($\overline{\omega}_{s}$) photons, respectively, into separate
output TMWs that exclude the pump, as shown in Fig.~\ref{CircuitII}.
This is achieved by bringing two auxiliary TMWs into proximity with
the principal TMW; all are of the same width $w_{1}$ so the spatial
profiles of both the even and odd modes can be maintained in the
course of coupling. The pump fails to couple to these auxiliary TMWs
because it has a short-range evanescent field by virtue of its high
frequency. The lengths of the auxiliary TMWs, and their separation
distances, are designed to optimize coupling of both the even and
the odd modes. For the low-frequency (high-frequency) coupler, the
auxiliary TMW is of length $L_{2}$ ($L_{3}$) and the separation
distance is $b_{1}$ ($b_{2}$) from the principal waveguide, as
illustrated in Fig.~\ref{CircuitII}(b).

\begin{figure}[t]
\centering
\includegraphics[width=3.4 in,totalheight =2.5 in]{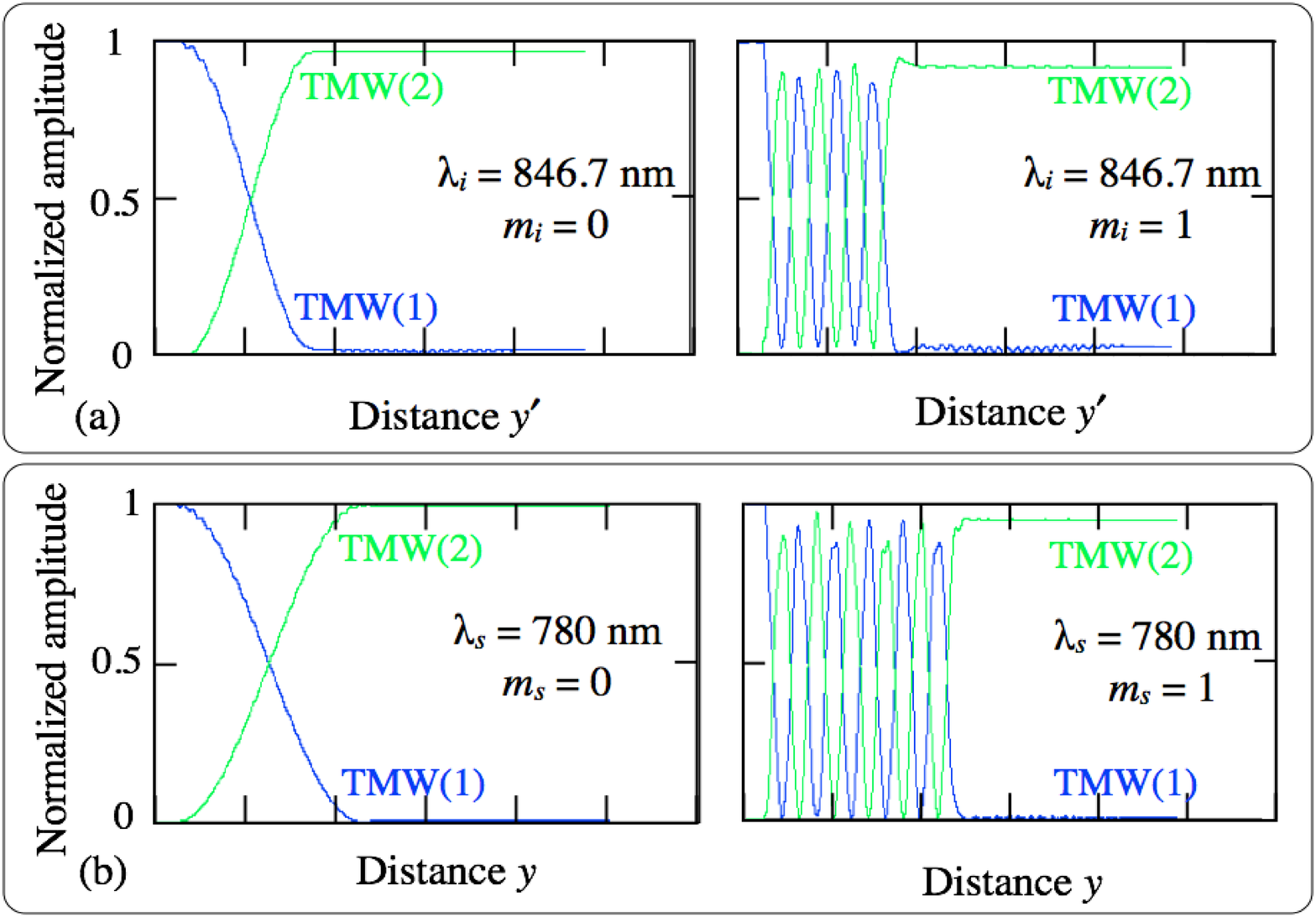}
\caption{Simulated performance of the photonic circuit described in
Fig.~\ref{CircuitII} for generating photons entangled in mode
number. The Type-0 $(e,e,e)$ SPDC source is identical to that used
in Fig.~\ref{ExSpectral}; nondegenerate photon pairs with center
wavelengths of 846.7 nm and 780 nm are generated using a poling
period of $\Lambda^{(1)} = 2.588 \: \mu$m. The circuit dimensions
are $w_{1}= 4.2 \:\mu$m, $L_{1}=2$ mm, $L_{2}=4.1 $ mm, $L_{3}=8.25$
mm, and $b_{1}= b_{2}= 4 \:\mu$m. The Bragg grating comprises two
multiplexed holograms, with periods of 83.8 and 83.7 nm, so both the
even and odd modes associated with the low-frequency photon can be
reflected. Panels (a) and (b) display the performance of the low-
and high-frequency couplers, respectively. Each panel contains
separate plots for the the fundamental ($m=0$) and first-order
($m=1$) modes. For the left (right) subplots, the blue and green
curves represent the evolution with distance of the normalized even
(odd) modes in the original and auxiliary TMWs, respectively.}
\label{ExModal}
\end{figure}

A simulation in which the photonic circuit in Fig.~\ref{CircuitII}
is used to generate a pair of photons entangled in mode number is
presented in Fig.~\ref{ExModal}. The nonlinear SPDC source is
identical to that displayed in Fig.~\ref{ExSpectral}. The Bragg
grating consists of two multiplexed holograms, with periods of 83.8
and 83.7 nm, to reflect the fundamental and first-order modes of the
low-frequency down-converted photon, respectively.
Figures~\ref{ExModal}(a) and \ref{ExModal}(b) display the
performance of the low- and high-frequency couplers, respectively.
Each panel has two subplots, for the even ($m=0$) and odd modes
($m=1$). For both frequencies, the coupling length of the
first-order mode is substantially shorter than that of the
fundamental mode since the peaks of the field for the former lie
closer to the transverse edges of the waveguide. Since the coupling
length is fixed, multiple cycles of energy transfer of the odd modes
(right panels) are required to assure full coupling of the even
modes (left panels).
\begin{figure}[t]
\centering
\includegraphics[width=3.4in,totalheight=5in]{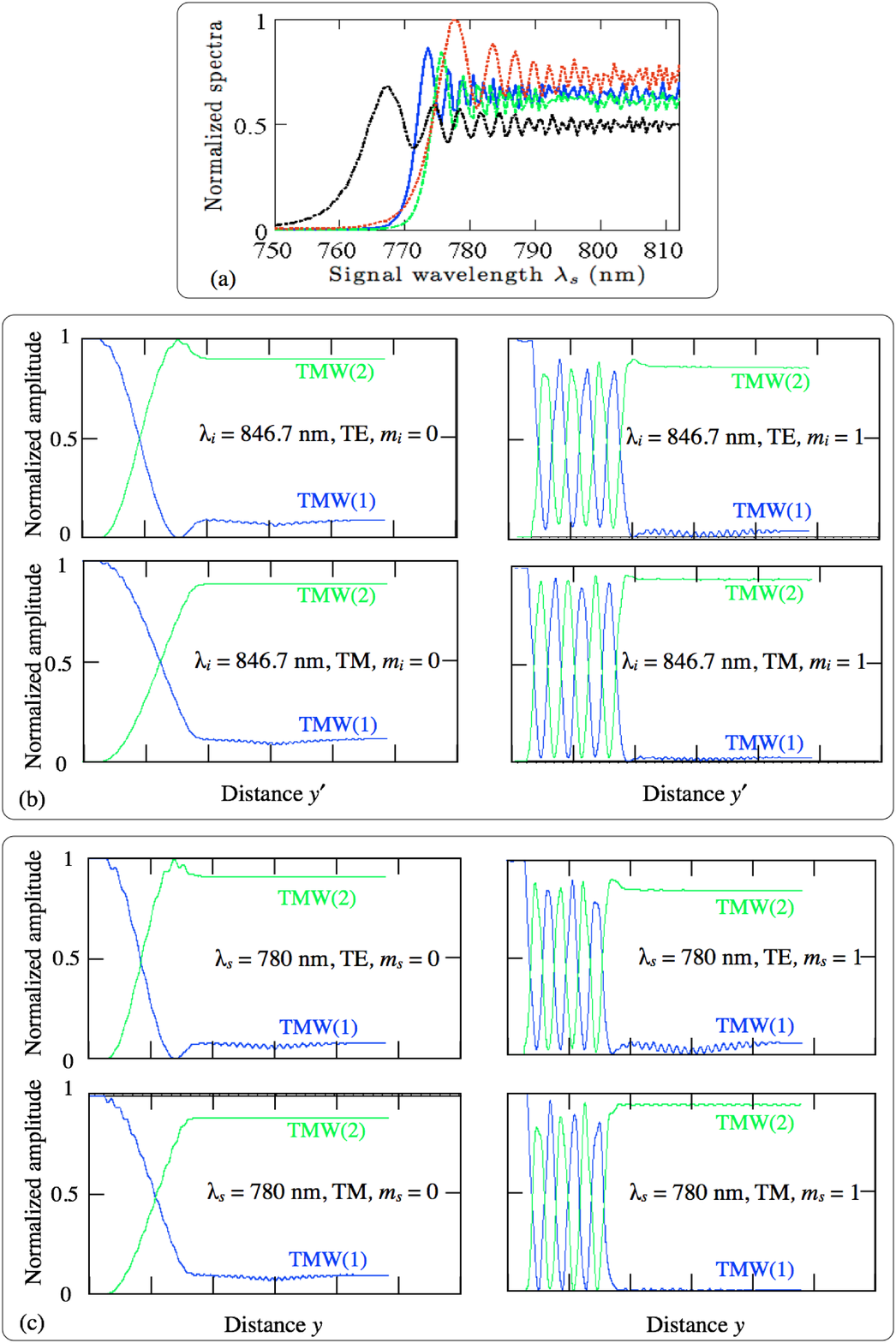}
\caption{Simulated performance of a modified version of the photonic
circuit displayed in Fig.~\ref{CircuitII} for generating photons
that are doubly entangled in mode number and polarization. An
$o$-polarized pump at $\lambda_{p} = 406$ nm, and in mode $m_{p}=1$,
is used to produce nondegenerate photons via Type-II $(o,e,o)$ and
$(e,o,o)$ SPDC processes. The preselected center wavelengths of the
down-converted photons are again 846.7 nm and 780 nm. The
periodically poled region is linearly chirped over a distance $L_{1}
= 20$ mm with an initial poling period of $1.84 \; \mu$m and a final
poling period of $1.87 \: \mu$m. The Bragg grating comprises four
multiplexed holograms, with periods 83.7, 83.8, 80.8 and 80.9 nm,
designed to reflect the low-frequency photons in the backward
direction. Other circuit dimensions are $w_{1}= 4.4 \: \mu$m,
$L_{2}=4.6 $ mm, $L_{3}=3.8$ mm, $b_{1}= 4 \: \mu$m, and $b_{2}= 3.5
\:\mu$m. Panel (a) shows the normalized spectra at the output of the
nonlinear region vs. the signal wavelength $\lambda_s$. The solid
blue and dashed green curves represent the output spectra associated
with Type-II $(o,e,o)$ SPDC, whereas the dash-dotted black and
dotted red curves represent the output spectra associated with
Type-II $(e,o,o)$ SPDC. Panels (b) and (c) display the performance
of the low- and high-frequency couplers, respectively. Each of these
panels contains four subplots: the top (bottom) subplots represent
TE (TM) polarization, while the left (right) subplots represent the
fundamental (first-order) mode. The blue and green curves represent
the evolution with distance of the normalized modes in the original
and auxiliary TMWs, respectively.} \label{ExDouble}
\end{figure}

A modified version of the photonic circuit displayed in
Fig.~\ref{CircuitII} can also be used to generate photons, via a
Type-II interaction, that are doubly entangled in mode number and
polarization. The biphoton state in that case is
\begin{equation}
\begin{array}{cl}
\vert \Psi\rangle\thicksim \displaystyle \int \mathrm{d}\omega_{s}&
\!\!\!\left[\Phi_{0,1,o,e}\left(\omega_{s}\right)\:\vert 0,o\rangle_{\omega_{s}}\vert 1,e\rangle_{\omega_{i}}+\:\Phi_{1,0,o,e}\left(\omega_{s}\right) \:\vert 1, o\rangle_{\omega_{s}}\vert 0,e\rangle_{\omega_{i}}\right. \\
 &\left.  \!\!\!+\:\Phi_{0,1,e,o}\left(\omega_{s}\right) \:\vert 0, e\rangle_{\omega_{s}}\vert 1,o\rangle_{\omega_{i}}+\:\Phi_{1,0,e,o}\left(\omega_{s}\right) \:\vert 1, e\rangle_{\omega_{s}}\vert 0,o\rangle_{\omega_{i}} \right].
\end{array}
\end{equation}
This interaction can be viewed as a superposition of Type-II
$(o,e,o)$ and Type-II $(e,o,o)$. To generate such doubly entangled
photons, the periodically poled nonlinear region portrayed in
Fig.~\ref{CircuitII} would have to be replaced by nonuniform (e.g.,
linearly chirped) poling, or aperiodic poling \cite{Norton04}, to
satisfy the phase matching conditions for the four possibilities.
Also, the holographic Bragg grating would then have to consist of
four multiplexed holograms of long length, or two multiplexed
holograms of short length. Since the Bragg-grating outputs in both
directions could be a photon in the fundamental- or first-order
mode, with TE- or TM-polarization, the low- and high-frequency
couplers would both have to be designed to optimize coupling for all
of the possibilities. It is worthy of mention that states entangled
in more than one degree of freedom are the primary resource for
one-way computation \cite{raussendorf01}.

An example to illustrate the simulated performance of such a
photonic circuit for generating photons that are doubly entangled in
mode number and polarization is displayed in Fig.~\ref{ExDouble}. As
shown in Fig.~\ref{ExDouble}(a), the use of linearly chirped poling
leads to broadband spectra. The low-frequency photon would be
filtered using the Bragg grating, while the high-frequency photon
could be filtered by placing a narrowband filter at the output
detector. The generation of narrowband spectra could, in principle,
be achieved by making use of aperiodic poling; however, this would
require poling periods so small that they might be beyond the
capabilities of current fabrication techniques.

\section{Photonic circuits for dispersion management}

The presence of dispersion is deleterious to the operation of
photonic circuits used for many quantum-information applications.
Dispersion results from the dependence of the propagation constant
$\beta$ on frequency, mode number, and polarization.
Polarization-mode dispersion generally provides the strongest
contribution, especially in a birefringent material such as
LiNbO$_{3}$. However, the Type-0 $(e,e,e)$ interaction, by virtue of
its three parallel electric field vectors, suffers only from
frequency and modal dispersion, and these are often sufficiently
weak that they can be neglected for short devices. This is not the
case for Type-II circuits, however, where polarization-mode
dispersion is indeed dominant.

In this section, we address the management of dispersion and
consider possible ways of mitigating it in photonic circuits,
whatever the type of interaction. The optimal approach for
circumventing dispersion depends on the particular application.

An example that proves instructive is the photonic-circuit HOM
interferometer, in which a simple adjustment of the path length can
be effective. The photonic circuit shown in Fig.~\ref{CircuitI}
provides for the generation, separation, and guiding of two SPDC
photons into two distinct SMW output ports. An HOM interferometer
can be fabricated on the same chip by feeding these outputs into a
3-dB coupler endowed with an electro-optic phase modulator in one of
its input arms. This allows a variable phase delay to be introduced
between the photons, thereby enabling an interferogram to be
generated.

The presence of dispersion alters the coincidence rate $R(\tau)$
associated with the HOM interferometer. The origin of this effect
can be understood by casting the interferometer as a four-port
system with each arm behaving as an ideal phase filter. Referring to
Fig.~\ref{CircuitI}, the matrix $\mathbf{T}$ that relates the output
and input of the system then takes the form
\begin{equation} \label{transmatrix}
\mathbf{T}=\left[ \begin{array}{cc}
H \left( \omega\right) & 0 \\
 0& H \left( \omega\right)\end{array}  \right] \, \left[ \begin{array}{cc}
1 & j \\
 j & 1\end{array} \right]\,\left[ \begin{array}{cc}
H\left( \omega\right)H_{o} \left( \omega\right) & 0 \\
0 &H\left( \omega\right)H_{e} \left( \omega\right)\end{array}  \right],
\end{equation}
with
\begin{equation}
\begin{array}{cl}
 H\left( \omega\right)&=\exp\left[ -j\beta_{1,\sigma}\left( \omega\right)\,l\right]  \\
 H_{o}\left( \omega\right)&=\exp\left[-j \beta_{1,\sigma}\left( \omega\right)\,l_{o}\right]  \\
 H_{e}\left( \omega\right)& =\exp\left[ -j\beta_{1,\sigma}\left( \omega\right)\,
 L_{t}/2
  -j\beta_{0,\sigma}\left( \omega\right)\left( l_{e}+ L_{t}/2\right) \right],
\end{array}
\label{phasefilters}
\end{equation}
where $l$ is the length of an arm of the 3-dB coupler; $l_{e}$ and
$l_{o}$ are the distances traveled by the even and odd modes from
the generation region to the 3-dB coupler, respectively; $L_{t}$ is
the length of the tapered region; and the sandwiched matrix
characterizes the lossless beam splitter. The mode in the
adiabatically tapered region is assumed to propagate with
propagation constant $\frac12 \left[ \beta_{0,\sigma}\left(
\omega\right)+\beta_{1,\sigma}\left( \omega\right)\right]$.

The presence of dispersion results in a modification of the
coincidence rate $R\left(\tau \right)$. In particular, the second
term of (\ref{coincidence}) is modified by the quantity
\begin{equation} \label{dispfactor}
D\left( \omega_{s}\right)=T_{11} ^{*}\left( \omega_{s}\right)T_{22}^{*} \left( \omega_{i}\right)T_{21} \left( \omega_{i}\right)T_{12} \left( \omega_{s}\right),
\end{equation}
where $\omega_{i}=\omega_{p}-\omega_{s}$, and the $T_{uv}$ are
elements of the $ \mathbf{T} $ matrix with $ u $ and $ v $
representing row and column indexes, respectively. Substituting
(\ref{transmatrix}) and (\ref{phasefilters}) into (\ref{dispfactor})
yields
\begin{equation}
D\left( \omega_{s}\right)=  \exp\left\lbrace j\left[ \beta_{1,\sigma_{s}}\left( \omega_s\right)-\beta_{1,\sigma_{i}}\left( \omega_i\right) \right] \left[l_{o}-L_{t}/2\right] \; -j\left[\beta_{0,\sigma_{s}}\left( \omega_s\right)-\beta_{0,\sigma_{i}}\left( \omega_i\right) \right] \left[l_{e}+L_{t}/2\right] \right\rbrace ,
\end{equation}
which, together with (\ref{eqPolingperiods}), gives rise to
\begin{equation}
D\left( \overline{\omega}_{s}\right)= \exp\left\lbrace j\left[
\beta_{1,\sigma_{s}}\left(\overline{\omega}_s\right)-\beta_{1,\sigma_{i}}
\left(\overline{\omega}_i\right) \right]
\left[l_{o}-l_{e}-L_{t}\right] \right. .
\end{equation}
It is apparent that choosing $l_{o}=l_{e}+L_{t}$ results in
dispersion cancelation at the preselected frequencies. Moreover, it
leads to even-order dispersion cancelation for other frequency components
as well.

\begin{figure}[t]
\centering
\includegraphics[width=3.4 in,totalheight =1.7 in]{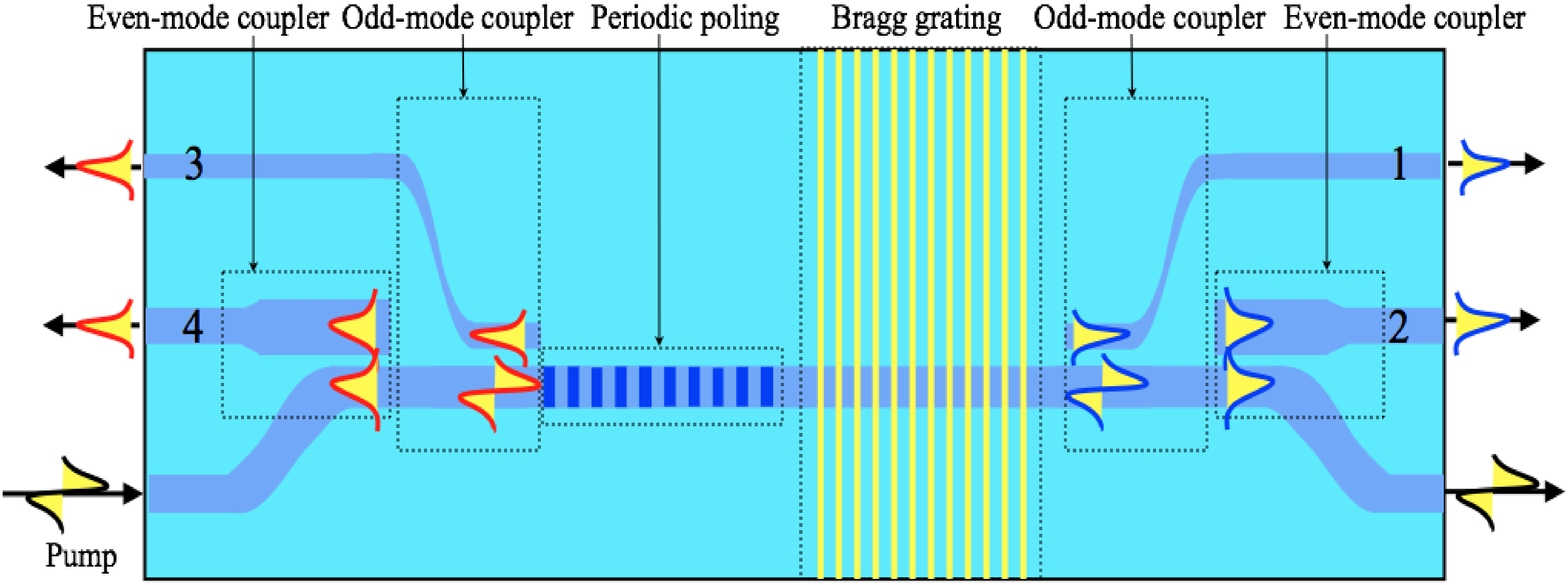}
\caption{Sketch of a photonic circuit that generates photons
entangled in path (not to scale). After generation in the nonlinear
region, they are separated first on the basis of their frequencies,
and then on the basis of their mode numbers.} \label{CircuitIII}
\end{figure}
More generally, the deleterious effects of dispersion can be
mitigated by mapping every possibility for each available degree of
freedom into a different path, resulting in the conversion of modal,
spectral, and polarization entanglement into path entanglement.

Consider, as an example, a Type-II $(e,o,o)$ interaction. In this
case, frequency and polarization are effectively only a single
degree of freedom since the photon with frequency above (below) the
degenerate value is always $e$ ($o$). Thus, two degrees of freedom
suffice for characterizing the system: frequency or polarization and
mode number. A photonic circuit for converting these two degrees of
freedom into photons entangled in path is sketched in
Fig.~\ref{CircuitIII}. The photons are first separated on the basis
of their frequencies, and then on the basis of their mode numbers.
Each output port thus carries only a particular degree of freedom,
thereby skirting the distinguishability engendered by
dispersion-induced time delay. The circuit displayed in
Fig.~\ref{CircuitIII} can be viewed as a superposition of those
illustrated in Figs.~\ref{CircuitI} and \ref{CircuitII}.

The biphoton state entangled in path can be written as
\begin{equation}
\vert \Psi\rangle\thicksim \displaystyle \int \:\mathrm{d}\omega_{s}\; \left[\Phi_{0,1,\sigma_{s},\sigma_{i}}\left(
\omega_{s}\right)\:\vert \omega_{s},0,\sigma_{s}\rangle_{1}\vert\omega_{i},1,\sigma_{i} \rangle_{3}+\:\Phi_{1,0,\sigma_{s},\sigma_{i}}\left( \omega_{s}\right) \:\vert\omega_{s},1, \sigma_{s}\rangle_{2}\vert\omega_{i},0,\sigma_{i}\rangle_{4}\right],
\end{equation}
where the subscripts 1, 2, 3, and 4 refer to the output ports
labeled in Fig.~\ref{CircuitIII}. The photon of frequency
$\omega_{i}$ is assumed to be back-reflected by the holographic
Bragg grating.

In the case of double entanglement, each output port in a circuit
such as that shown in Fig.~\ref{CircuitIII} would deliver photons
with either TE- or TM-polarization. To avoid distinguishability as a
result of polarization-mode dispersion, two methods could be used:
1) each output port could be split into a pair of ports, resulting
in eight doubly entangled paths; or 2) an electro-optic
TE$\rightleftharpoons$TM mode converter \cite{Alferness80} could be
used at the path center of each output port, thereby advancing the
slow, and retarding the fast, photons in such a way that they would
arrive simultaneously at the output detectors.

\section{Conclusion}
We have presented several photonic-circuit designs based on
Ti:LiNbO$_{3}$ diffused channel waveguides for generating photons
with various combinations of modal, spectral, and polarization
entanglement. These circuits contain, in various forms, a nonlinear
periodically poled two-mode waveguide structure, single- and
two-mode waveguide-based couplers, and multiplexed holographic Bragg
gratings. Depending on the choice of parameters, the first photonic
circuit produces frequency-degenerate down-converted photons with
even spatial parity, nondegenerate down-converted photons entangled
in frequency, or nondegenerate down-converted photons simultaneously
entangled in frequency and polarization. The second photonic circuit
produces modal entanglement, or photons that are doubly entangled in
mode number and polarization. The third photonic circuit converts
modal, spectral, and polarization entanglement into path
entanglement to mitigate the effects of dispersion. Simulations have
been carried out, with the help of the the commercial photonic and
network design software package RSoft, to gauge the performance of
these circuits. Furthermore, since Ti:LiNbO$_{3}$ photonic circuits
have the salutary property of permitting the generation,
transmission, and processing of photons to be accommodated on a
single chip, we expect that they will find use in the efficient
implementation of various quantum information processing schemes
\cite{saleh10b}.


\begin{thebibliography}{10}
\providecommand{\url}[1]{#1}
\csname url@samestyle\endcsname
\providecommand{\newblock}{\relax}
\providecommand{\bibinfo}[2]{#2}
\providecommand{\BIBentrySTDinterwordspacing}{\spaceskip=0pt\relax}
\providecommand{\BIBentryALTinterwordstretchfactor}{4}
\providecommand{\BIBentryALTinterwordspacing}{\spaceskip=\fontdimen2\font plus
\BIBentryALTinterwordstretchfactor\fontdimen3\font minus
  \fontdimen4\font\relax}
\providecommand{\BIBforeignlanguage}[2]{{%
\expandafter\ifx\csname l@#1\endcsname\relax
\typeout{** WARNING: IEEEtran.bst: No hyphenation pattern has been}%
\typeout{** loaded for the language `#1'. Using the pattern for}%
\typeout{** the default language instead.}%
\else
\language=\csname l@#1\endcsname
\fi
#2}}
\providecommand{\BIBdecl}{\relax}
\BIBdecl

\bibitem{bennett84}
C.~H. Bennett and G.~Brassard, ``Quantum cryptography: {P}ublic key
  distribution and coin tossing,'' in \emph{Proceedings of the International
  Conference on Computers, Systems \& Signal Processing}.\hskip 1em plus 0.5em
  minus 0.4em\relax Bangalore, India: Institute of Electrical and Electronics
  Engineers, Dec. 1984, pp. 175--179.

\bibitem{klyshko80}
D.~N. Klyshko, \emph{Photons and Nonlinear Optics}.\hskip 1em plus 0.5em minus
  0.4em\relax Moscow: Nauka, 1980, {C}haps. 1 and 6. Translation: Gordon and
  Breach, New York, 1988.

\bibitem{joobeur94}
A.~Joobeur, B.~E.~A. Saleh, and M.~C. Teich, ``Spatiotemporal coherence
  properties of entangled light beams generated by parametric
  down-conversion,'' \emph{Phys. Rev. A}, vol.~50, pp. 3349--3361, Oct. 1994.

\bibitem{joobeur96}
A.~Joobeur, B.~E.~A. Saleh, T.~S. Larchuk, and M.~C. Teich, ``Coherence
  properties of entangled light beams generated by parametric downconversion:
  {T}heory and experiment,'' \emph{Phys. Rev. A}, vol.~53, pp. 4360--4371, Jun.
  1996.

\bibitem{saleh00}
B.~E.~A. Saleh, A.~F. Abouraddy, A.~V. Sergienko, and M.~C. Teich, ``Duality
  between partial coherence and partial entanglement,'' \emph{Phys. Rev. A},
  vol.~62, pp. 043\,816, Sep. 2000.

\bibitem{rarity90}
J.~G. Rarity and P.~R. Tapster, ``Experimental violation of {B}ell's inequality
  based on phase and momentum,'' \emph{Phys. Rev. Lett.}, vol.~64, pp.
  2495--2498, May 1990.

\bibitem{franson99}
J.~D. Franson, ``Inconsistency of local realistic descriptions of two-photon
  interferometer experiments,'' \emph{Phys. Rev. A}, vol.~61, p. 012105, Dec.
  1999.

\bibitem{mair01}
A.~Mair, A.~Vaziri, G.~Weihs, and A.~Zeilinger, ``Entanglement of the orbital
  angular momentum states of photons,'' \emph{Nature}, vol. 412, pp. 313--316,
  Jul. 2001.

\bibitem{Walborn05}
S.~P. Walborn, S.~P{\'a}dua, and C.~H. Monken, ``Conservation and entanglement
  of {H}ermite--{G}aussian modes in parametric down-conversion,'' \emph{Phys.
  Rev. A}, vol.~71, pp. 053\,812, May 2005.

\bibitem{abouraddy07}
A.~F. Abouraddy, T.~Yarnall, B.~E.~A. Saleh, and M.~C. Teich, ``Violation of
  {B}ell's inequality with continuous spatial variables,'' \emph{Phys. Rev. A},
  vol.~75, pp. 052\,114, May 2007.

\bibitem{yarnall07a}
T.~Yarnall, A.~F. Abouraddy, B.~E.~A. Saleh, and M.~C. Teich, ``Experimental
  violation of {B}ell's inequality in spatial-parity space,'' \emph{Phys. Rev.
  Lett.}, vol.~99, pp. 170\,408, Oct. 2007.

\bibitem{yarnall07b}
------, ``Synthesis and analysis of entangled photonic qubits in spatial-parity
  space,'' \emph{Phys. Rev. Lett.}, vol.~99, pp. 250\,502, Dec. 2007.

\bibitem{Saleh09}
M.~F. Saleh, B.~E.~A. Saleh, and M.~C. Teich, ``Modal, spectral, and
  polarization entanglement in guided-wave parametric down-conversion,''
  \emph{Phys. Rev. A}, vol.~79, pp. 053\,842, May 2009.

\bibitem{saleh10b}
M.~F. Saleh, G.~{Di~Giuseppe}, B.~E.~A. Saleh, and M.~C. Teich, ``Modal and
  polarization qubits in {Ti:LiNbO$_3$} photonic circuits for a universal
  quantum logic gate,'' {http://arxiv.org/abs/1007.3256}. {S}ubmitted for
  publication, Jul. 2010.

\bibitem{raussendorf01}
R.~Raussendorf and H.~J. Briegel, ``A one-way quantum computer,'' \emph{Phys.
  Rev. Lett.}, vol.~86, pp. 5188--5191, May 2001.

\bibitem{knill01}
E.~Knill, R.~Laflamme, and G.~J. Milburn, ``A scheme for efficient quantum
  computation with linear optics,'' \emph{Nature}, vol. 409, pp. 46--52, Jan.
  2001.

\bibitem{Politi08}
A.~Politi, M.~J. Cryan, J.~G. Rarity, S.~Yu, and J.~L. {O'Brien},
  ``Silica-on-silicon waveguide quantum circuits,'' \emph{Science}, vol. 320,
  pp. 646--649, May 2008.

\bibitem{Matthews09}
J.~C.~F. Matthews, A.~Politi, A.~Stefanov, and J.~L. {O'Brien}, ``Manipulation
  of multiphoton entanglement in waveguide quantum circuits,'' \emph{Nature
  Photon.}, vol.~3, pp. 346--350, Jun. 2009.

\bibitem{Bennett98}
C.~H. Bennett and P.~W. Shor, ``Quantum information theory,'' \emph{{IEEE}
  Trans. Inform. Theory}, vol.~44, pp. 2724--2742, Oct. 1998.

\bibitem{Nielsen00}
M.~A. Nielsen and I.~L. Chuang, \emph{Quantum Computation and Quantum
  Information}.\hskip 1em plus 0.5em minus 0.4em\relax Cambridge, UK: Cambridge
  University Press, 2000.

\bibitem{OBrien09}
J.~L. {O'Brien}, A.~Furusawa, and J.~Vu{\v c}kovi{\'c}, ``Photonic quantum
  technologies,'' \emph{Nature Photon.}, vol.~3, pp. 687--695, Dec. 2009.

\bibitem{Politi09}
A.~Politi, J.~C.~F. Matthews, M.~G. Thompson, and J.~L. {O'Brien}, ``Integrated
  quantum photonics,'' \emph{{IEEE} J. Sel. Topics Quantum Electron.}, vol.~15,
  pp. 1673--1684, Dec. 2009.

\bibitem{Cincotti09}
G.~Cincotti, ``Prospects on planar quantum computing,'' \emph{J. Lightwave
  Technol.}, vol.~27, pp. 5755--5766, Dec. 2009.

\bibitem{ladd10}
T.~D. Ladd, F.~Jelezko, R.~Laflamme, Y.~Nakamura, C.~Monroe, and J.~L.
  {O'Brien}, ``Quantum computers,'' \emph{Nature}, vol. 464, pp. 45--53, Mar.
  2010.

\bibitem{fiorentino07}
M.~Fiorentino, S.~M. Spillane, R.~G. Beausoleil, T.~D. Roberts, P.~Battle, and
  M.~W. Munro, ``Spontaneous parametric down-conversion in periodically poled
  {KTP} waveguides and bulk crystals,'' \emph{Opt. Express}, vol.~15, pp.
  7479--7488, Jun. 2007.

\bibitem{Avenhaus09}
M.~Avenhaus, M.~V. Chekhova, L.~A. Krivitsky, G.~Leuchs, and C.~Silberhorn,
  ``Experimental verification of high spectral entanglement for pulsed
  waveguided spontaneous parametric down-conversion,'' \emph{Phys. Rev. A},
  vol.~79, pp. 043\,836, Apr. 2009.

\bibitem{Mosley09}
P.~J. Mosley, A.~Christ, A.~Eckstein, and C.~Silberhorn, ``Direct measurement
  of the spatial-spectral structure of waveguided parametric down-conversion,''
  \emph{Phys. Rev. Lett.}, vol. 103, pp. 233\,901, Dec. 2009.

\bibitem{zhong09}
T.~Zhong, F.~N. Wong, T.~D. Roberts, and P.~Battle, ``High performance
  photon-pair source based on a fiber coupled periodically poled {KTiOPO$_4$}
  waveguide,'' \emph{Opt. Express}, vol.~17, pp. 12\,019--12\,030, Jul. 2009.

\bibitem{chen09}
J.~Chen, A.~J. Pearlman, A.~Ling, J.~Fan, and A.~Migdall, ``A versatile
  waveguide source of photon pairs for chip-scale quantum information
  processing,'' \emph{Opt. Express}, vol.~17, pp. 6\,727--6\,740, Apr. 2009.

\bibitem{tanzilli01}
S.~Tanzilli, H.~{De~Riedmatten}, W.~Tittel, H.~Zbinden, P.~Baldi,
  M.~{De~Micheli}, D.~B. Ostrowsky, and N.~Gisin, ``Highly efficient
  photon-pair source using periodically poled lithium niobate waveguide,''
  \emph{Electron. Lett.}, vol.~37, pp. 26--28, Jan. 2001.

\bibitem{booth02}
M.~C. Booth, M.~Atat{\"u}re, G.~{Di~Giuseppe}, B.~E.~A. Saleh, A.~V. Sergienko,
  and M.~C. Teich, ``Counterpropagating entangled photons from a waveguide with
  periodic nonlinearity,'' \emph{Phys. Rev. A}, vol.~66, pp. 023\,815, Aug.
  2002.

\bibitem{dechatellus06}
H.~{Guillet~de~Chatellus}, A.~V. Sergienko, B.~E.~A. Saleh, M.~C. Teich, and
  G.~{Di~Giuseppe}, ``Non-collinear and non-degenerate polarization-entangled
  photon generation via concurrent type-{I} parametric downconversion in
  {PPLN},'' \emph{Opt. Express}, vol.~14, pp. 10\,060--10\,072, Oct. 2006.

\bibitem{Nishihara89}
H.~Nishihara, M.~Haruna, and T.~Suhara, \emph{Optical Integrated
  Circuits}.\hskip 1em plus 0.5em minus 0.4em\relax New York: McGraw--Hill,
  1989.

\bibitem{Saleh07}
B.~E.~A. Saleh and M.~C. Teich, \emph{Fundamentals of Photonics}, 2nd~ed.\hskip
  1em plus 0.5em minus 0.4em\relax Hoboken, NJ: Wiley, 2007.

\bibitem{Busacca04}
A.~C. Busacca, C.~L. Sones, R.~W. Eason, and S.~Mailis, ``First-order
  quasi-phase-matched blue light generation in surface-poled {Ti:indiffused}
  lithium niobate waveguides,'' \emph{Appl. Phys. Lett.}, vol.~84, pp.
  4430--4432, May 2004.

\bibitem{Lee04}
Y.~L. Lee, C.~Jung, Y.-C. Noh, M.~Park, C.~Byeon, D.-K. Ko, and J.~Lee,
  ``Channel-selective wavelength conversion and tuning in periodically poled
  {Ti:LiNbO$_3$} waveguides,'' \emph{Opt. Express}, vol.~12, pp. 2649--2655,
  Jun. 2004.

\bibitem{Alferness78}
R.~C. Alferness and R.~V. Schmidt, ``Tunable optical waveguide directional
  coupler filter,'' \emph{Appl. Phys. Lett.}, vol.~33, pp. 161--163, Jul. 1978.

\bibitem{Alferness80}
R.~C. Alferness, ``Efficient waveguide electro-optic {TE$\rightleftharpoons$TM}
  mode converter/wavelength filter,'' \emph{Appl. Phys. Lett.}, vol.~36, pp.
  513--515, Apr. 1980.

\bibitem{Hukriede03}
J.~Hukriede, D.~Runde, and D.~Kip, ``Fabrication and application of holographic
  {B}ragg gratings in lithium niobate channel waveguides,'' \emph{J. Phys. D:
  Appl. Phys.}, vol.~36, pp. R1--R16, Jan. 2003.

\bibitem{Runde07}
D.~Runde, S.~Brunken, S.~Breuer, and D.~Kip, ``Integrated-optical add/drop
  multiplexer for {DWDM} in lithium niobate,'' \emph{Appl. Phys. B: Lasers
  Opt.}, vol.~88, pp. 83--88, Jun. 2007.

\bibitem{Runde08}
D.~Runde, S.~Breuer, and D.~Kip, ``Mode-selective coupler for wavelength
  multiplexing using {LiNbO$_3$:Ti} optical waveguides,'' \emph{Cent. Eur. J.
  Phys.}, vol.~6, pp. 588--592, Sep. 2008.

\bibitem{schmidt76}
R.~V. Schmidt and H.~Kogelnik, ``Electro-optically switched coupler with
  stepped {$\Delta\beta$} reversal using {Ti-diffused} {LiNbO$_3$}
  waveguides,'' \emph{Appl. Phys. Lett.}, vol.~28, pp. 503--506, May 1976.

\bibitem{kogelnik76}
H.~Kogelnik and R.~V. Schmidt, ``Switched directional couplers with alternating
  {$\Delta\beta$},'' \emph{IEEE J. Quantum Electron.}, vol. QE-12, pp.
  396--401, Jul. 1976.

\bibitem{Fejer92}
M.~M. Fejer, G.~A. Magel, D.~H. Jundt, and R.~L. Byer, ``Quasi-phase-matched
  second harmonic generation: {T}uning and tolerances,'' \emph{IEEE J. Quantum
  Electron.}, vol.~28, pp. 2631--2654, Nov. 1992.

\bibitem{Hong87}
C.~K. Hong, Z.~Y. Ou, and L.~Mandel, ``Measurement of subpicosecond time
  intervals between two photons by interference,'' \emph{Phys. Rev. Lett.},
  vol.~59, pp. 2044--2046, Nov. 1987.

\bibitem{Campos90}
R.~A. Campos, B.~E.~A. Saleh, and M.~C. Teich, ``Fourth-order interference of
  joint single-photon wave packets in lossless optical systems,'' \emph{Phys.
  Rev. A}, vol.~42, pp. 4127--4137, Oct. 1990.

\bibitem{jundt97}
D.~H. Jundt, ``Temperature-dependent {S}ellmeier equation for the index of
  refraction, $n_{\rm e}$, in congruent lithium niobate,'' \emph{Opt. Lett.},
  vol.~22, pp. 1553--1555, Oct. 1997.

\bibitem{Wong02}
K.~K. Wong, Ed., \emph{Properties of Lithium Niobate}.\hskip 1em plus 0.5em
  minus 0.4em\relax Stevenage, U.K.: Institution of Electrical Engineers, 2002.

\bibitem{Feit83}
M.~D. Feit, J.~A. {Fleck, Jr.}, and L.~McCaughan, ``Comparison of calculated
  and measured performance of diffused channel-waveguide couplers,'' \emph{J.
  Opt. Soc. Am.}, vol.~73, pp. 1296--1304, Oct. 1983.

\bibitem{Hutcheson87}
S.~K. Korotky and R.~C. Alferness, ``{Ti:LiNbO$_3$} integrated optic
  technology,'' in \emph{Integrated Optical Circuits and Components: Design and
  Applications}, L.~D. Hutcheson, Ed.\hskip 1em plus 0.5em minus 0.4em\relax
  New York: Marcel Dekker, 1987.

\bibitem{Hocker77}
G.~B. Hocker and W.~K. Burns, ``Mode dispersion in diffused channel waveguides
  by the effective index method,'' \emph{Appl. Opt.}, vol.~16, pp. 113--118,
  Jan. 1977.

\bibitem{Gisin02}
N.~Gisin, G.~Ribordy, W.~Tittel, and H.~Zbinden, ``Quantum cryptography,''
  \emph{Rev. Mod. Phys.}, vol.~74, pp. 145--195, Mar. 2002.

\bibitem{Myers95}
L.~E. Myers, R.~C. Eckardt, M.~M. Fejer, R.~L. Byer, W.~R. Bosenberg, and J.~W.
  Pierce, ``Quasi-phase-matched optical parametric oscillators in bulk
  periodically poled {LiNbO$_3$},'' \emph{J. Opt. Soc. Am. B}, vol.~12, pp.
  2102--2116, Nov. 1995.

\bibitem{Ferriere06}
R.~Ferriere, B.-E. Benkelfat, J.~M. Dudley, and K.~Ghoumid, ``Bragg mirror
  inscription on {LiNbO$_3$} waveguides by index microstructuration,''
  \emph{Appl. Opt.}, vol.~45, pp. 3553--3560, May 2006.

\bibitem{Norton04}
A.~Norton and C.~de~{S}terke, ``Aperiodic {1}-dimensional structures for
  quasi-phase matching,'' \emph{Opt. Express}, vol.~12, pp. 841--846, Mar.
  2004.

\end{thebibliography}


\end{document}